\documentclass[12pt,a4paper,final]{iopart}

\usepackage{iopams}
\usepackage{graphicx}
\usepackage[breaklinks=true,colorlinks=true,linkcolor=blue,urlcolor=blue,citecolor=blue]{hyperref}
\usepackage{siunitx}
\expandafter\let\csname equation*\endcsname\relax
\expandafter\let\csname endequation*\endcsname\relax
\usepackage{amsmath}

\newcommand{\ts}{\textsuperscript}
\def\cotan{\qopname\relax o{cotan}}
\def\cosec{\qopname\relax o{csc}}

\begin{document}

\title{The International Young Physicists' Tournament 2017}

\author[cor1]{Martin Plesch$^{1,2}$}
\address{$^1$Institute of Physics, Slovak Academy of Sciences, Bratislava, Slovakia}
\address{$^2$Institute of Computer Science, Masaryk University, Brno, Czech Republic}
\ead{martin.plesch@savba.sk}

\author{Matej Badin}
\address{Faculty of Mathematics, Physics and Informatics, Comenius University, Bratislava, Slovakia}

\author{Nat\'alia Ru\v zi\v ckov\'a}
\address{Faculty of Mathematics, Physics and Informatics, Comenius University, Bratislava, Slovakia}

\begin{abstract}
In July 2017, National University of Singapore hosted the 30\ts{th} International Young Physicists' Tournament (IYPT) with symbolic participation of $30$ competing teams. IYPT is a modern scientific competition for teams of high school students, also known as Physics World Cup. Here we present a brief report from this years competition with a concise list of problems solved by participating teams. For one of the problems, Torsion gyroscope, we also bring a glimpse into its theoretical and experimental solution.

\end{abstract}

\vspace{2pc}

\section{Introduction}

After six days of gripping physics fights and entertaining cultural activities, 30\ts{th} IYPT ended on July 11\ts{th} 2017 in Singapore. The event, hosted by the National University of Singapore, paid a visit to the country which dominates this competition since several years. A combination of smart students, brilliant preparation, experience and English as mother tongue made the hosting team, yet again, the overall winner. Teams from China, Poland and Hungary made a very good sparring for Singapore in the final fight and well deserved their gold medals as well.

International Young Physicists' Tournament \cite{iypt.org, IYPT2016} is a modern competition not only because of its emphasis on teamwork, but also because of its complexity and form, which encourages students to work and present solutions similar to manner of senior scientists.
Each year in summer, $17$ scientifically interesting open-ended problems are announced and students are encouraged to work on their original solutions for as long as the whole school year. Their research on majority of tasks includes studying of existing literature and scientific work already devoted to the phenomenon. Students are asked to explain the physical principles behind the phenomenon, construct and perform their own experiments and analyse results of their investigation, as well as approach the problem theoretically. As IYPT problems are open-ended, they have non-trivial solutions and are often subject to current research, the whole process of preparation includes cooperation with scientists working in various fields of physics and natural sciences in general.

After finalising their investigation, team members prepare a $12$ minute presentation summarising their results, theoretical models and conclusions for each of the problems. These are then presented, discussed and opposed by other teams and the whole performance is judged by an independent jury consisting of scientists and physics teachers. In the next section we will briefly introduce the course of IYPT and some of the $17$ problems of last year. We will investigate a bit more in detail one of the problems, Torsion gyroscope, in section \ref{section-gyroscope} and conclude in the last section.

\section {30\ts{th} IYPT 2017}

\subsection{Course of the Tournament}

Throughout the competition, students experience five fights (each taking about half a day) and brakes between are filled with excursions and free time activities. The same basically holds for jurors, except for the first day when they have to attend a jury meeting aiming to make the grading as fair as possible towards all participants.

This year, the host of IYPT was National University of Singapore (NUS), ranking among the top $25$ universities in the world. The event started on 6\ts{th} of July with an opening ceremony in the morning and already the first physics fight in the afternoon.
The final fight and awards ceremony were held on July 10\ts{th}, followed by a dinner party, where teams could celebrate and enjoy spending time with their new international friends. The final day was no less entertaining: a visit of Universal Studios Amusement Park had been a nice ending of the whole Tournament for students -- their leaders still had to work to prepare a new set of problems for the next year.

With $30$ participating teams, 30\ts{th} IYPT broke the participation record once again. During the last $3$ years, $5$ new countries joined the IYPT, making it a fast growing event. This in spite of the fact that the very high scientific level of presented solutions and individual performances of students combined with tournament's complex format presents an entrance barrier for new teams. To help perspective teams to gain experience before fully joining the competition, a new concept of Guest teams has been introduced and tested in IYPT 2017. The first ever guest team to participate was the team from India, which struggled a bit to get a full scale preparation support. Three motivated students came to Singapore to watch other teams competing, learn about the course of the Tournament, its traditions and level of solutions and discussions. They even had a chance to challenge themselves and participate in a friendly physics fight with team Singapore B. Students found the possibility of presenting results of their own research and discuss them in front of an IYPT jury as extremely useful and motivating and are willing to take part next year as a fully competing team.

Within IYPT, each team takes part in five selective fights - in each fight, three, occasionally four teams meet and discuss their solutions. Grades given by juries for each performance are summed up and lead to a ranking table that determines the composition of the final. Based on absolute ranking, first three teams automatically proceed; optionally, the highest ranked team of those winning all their selective fights is the fourth final participant. So far, winning all selective fights and not heading directly to the finals was rather rare. With growing number of teams, the situation changes: in Singapore, both Hungary and New Zealand have won all their selective fights and yet New Zealand did not became a finalist, as four teams is a maximum.

This and few other aspects from IYPT 2017 showed the need of a serious consideration of the question how a further growth of the Tournament deteriorates the adequacy of current tournament system. As usually three teams meet in a single physics fight, a team meets only $10$ other teams during the competition. With $30$ teams present it becomes a small fraction and the probability of the best teams meeting directly in one fight decreases with the number of participants. One option for dealing with growing tournament is increasing the number of selective fights. However, besides being it a logistical problem, five physics fights and the finals is already stressful and exhausting for the participants. E.g. in Physics Olympiad \cite{IPhO} students of the same age compete during roughly two half-days, in IJSO \cite{IJSO} three half-days, in IYPT students compete for five to six half-days already. This is why one would have to think about concepts including semifinals, power-pairing or deploying based on previous results.

Getting to know the history and culture of the host country is another important aspect of IYPT, and this year it was not different. Participants had a chance to take part in a guided tour of Singapore where they learned about the foundation of the country, its diverse culture as well as about landscape of modern Singapore and precise architectural planning. Meanwhile, other group of students and their leaders went to explore laboratories of the Physics department of the NUS, among them Astrophysics observatory, Centre of quantum technologies and Biophysics laboratory and found out about their world-leading research. Apart from that, there was a half day excursion to popular Gardens by the Bay ending with a light show at night. This all made IYPT 2017 a memorable event for each participant.

\subsection{Problems}
As each year, $17$ interesting open-ended physics problems were published nearly a year in advance before IYPT 2017 \cite{problems}.

The procedure of choosing IYPT problems is complex: problems are based on ideas collected from people all over the world via internet submission system, where anyone is welcome to submit proposals anytime. From this collection, IYPT Problem Committee chooses a subset of suitable ones to be voted on by the International Organising Committee of IYPT. Set of problems should contain topics from all fields of high school physics including mechanics, electricity and magnetism, optics, fluid dynamics or thermodynamics. To some problems also knowledge from other science fields like computer science or chemistry can be helpful. All problems should allow experimental realisation and at least qualitative theoretical approach. Optimally, problems have several levels of possible solutions, allowing teams to compete from regional levels within participating countries up to the final of IYPT. This makes the selection and precise formulation of IYPT problems a very hard task.

Among IYPT 2017 problems, visually impressive were Schlieren photography experiments, which were the core of Problem 8. \textit{Visualising density}. Schlieren photography is based on detecting change of path of light due to density gradient in the medium it passes trough. That can be done by focusing light with a mirror and inserting a blade to the point of focus. The rays which have changed their path will be stopped by the blade and that is seen as difference in intensity. The other option is to use rainbow colour filter instead of a blade. To fulfil the task, teams had to build their own Schlieren setup and investigate resolution of their apparatus.

Mostly mechanics problems were tasks No. 6. \textit{Falling chain}, 12. \textit{Torsion gyroscope}, 14. \textit{Gee-Haw Whammy Diddle} and 16. \textit{Metronome synchronisation}.
Striking was the Falling chain problem, which required students to explain why a ladder-like chain of wooden sticks can, when landing on a hard surface, fall faster  than objects in a free fall. The trick was hidden in conservation of angular momentum after impact of a stick, which in the end fastened the whole chain. Unexpectedly, it was very difficult to perform measurements accurate enough to observe the phenomenon and analyse them with sufficiently small errors.
The idea of task 14 originated in observing a scientific toy \textit{consisting of a simple wooden stick and a second stick that is made up of a series of notches with a propeller at its end. When the wooden stick is pulled over the notches, the propeller starts to rotate}.
Explaining why and how the propeller rotates turned out to be more difficult and complex than expected and inspired even a scientific publication \cite{vrtulka} in cooperation of students, their leaders and jurors that significantly advanced knowledge on the phenomenon.

Another toy to build and investigate was the No. 2. \textit{Balloon Airhorn}. Surprising was the fact that the nature of sound of such airhorn depended on the combination of balloon parameters and tube length used for the horn. If the frequency dictated by the membrane stiffness and the overpressure in the cup caused by blowing was similar to resonant frequencies of the tube, the system behaved like a half-tube resonator. However, if the tube length was too small, the airhorn resembled a spring-mass system, with air in the tube acting as the membrane load. Due to this ambiguity and strong parameter dependence, it was rather difficult for teams to defend their solutions during discussions and inevitably led to large spread in jury grading.

No less challenging was problem No. 4. \textit{Magnetic Hills}, the only problem from magnetism at IYPT 2017, which was about investigating hill-like structures which arise after immersing ferrofluid into inhomogeneous magnetic field. Hill-like structure arising from initial instability depended on magnetic field strength and pattern, liquid susceptibility and also surface tension and density. As part of their solution, some teams even mixed their own ferrofluid.

Eggs were a theme of two 2017 problems: one of them was problem No. 15, \textit{Boiled egg}. Participants were to come up with \textit{non-invasive methods to detect the degree to which a hen's egg is cooked by boiling} and investigate the sensitivity of their methods. A well known method is to twist the egg, then stop the rotation by slightly touching it with a finger. If the egg is raw, liquid yolk inside is still rotating and thus the whole egg continues to rotate. As soon as most of the egg is solidified by cooking, the whole egg is stopped by the touch.
One of the methods used by IYPT teams was based on a similar principle: immediately after releasing it on an inclined plane, an uncooked egg has bigger angular acceleration than a cooked one, because of the non-rotating liquid yolk. That results in lower effective moment of inertia compared to a solid cooked egg. However, this method turned out not to be sensitive enough. On the other hand, one of the most sensitive methods was based on analysing energy losses in twisting eggs. Eggs were made to oscillate torsionally and the damping coefficient of these oscillations could quite precisely determine how well cooked the egg was. The weaker the damping, the longer the cooking time.

The first task involved experiments with eggs as well. Traditionally, it was an \textit{Invent yourself} problem, which often requires participants to design and build a unique device with specified functions. This year, the task was to \textit{Construct a passive device that will provide safe landing
for an uncooked hen's egg} dropped from 2.5 meters. The special requirement was to minimise the volume of such device.

As usually, doing experiments to problem No. 17 was fun: \textit{A 'vacuum bazooka' can be built with a simple plastic pipe, a light projectile, and a vacuum cleaner.} Participants were supposed to maximise nozzle velocity of the projectile. We have seen many experiments crashing cans, playing bricks and other material, but also a lot of deep physics of fluid dynamics, (under)-caliber projectiles and vacuum-cleaner technology.

\section {Physics behind Torsion gyroscope}
\label{section-gyroscope}
\begin{figure}[ht]
\centering
  \includegraphics[width=0.4\textwidth]{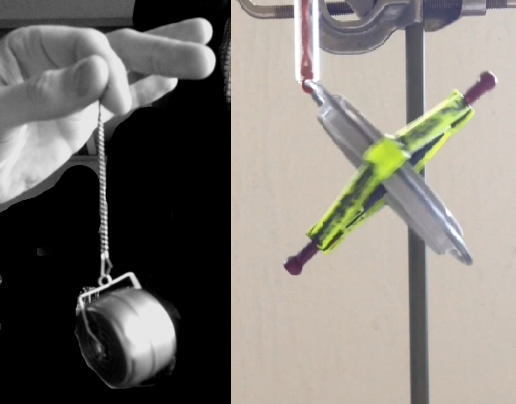}
\caption{Left: Figure from the original problem statement. Rotating wheel attached to a vertical thread. Right: Our realisation of the torsion gyroscope. Toy gyroscope attached to a rubber thread which lies inside a plastic tube, in order to suppress any motion of the thread except the rotation around its own axis. }
\label{fig_problem}
\end{figure}

Here we provide a brief glimpse into the physics behind the IYPT problem No. 12. \textit{Torsion gyroscope}. The problem states: \textit{Fasten the axis of a wheel to a vertical thread that has a certain torsional resistance (see Figure \ref{fig_problem}). Twist the thread, spin the wheel, and release it. Investigate the dynamics of this system.}

In the following lines we briefly explain the nature of the phenomenon, provide a minimal model which is able to describe the dynamics of the system, as well as full equations of motion which cover all subtleties of the motion.

\begin{figure}[ht]
\centering
  \includegraphics[width=0.5\textwidth]{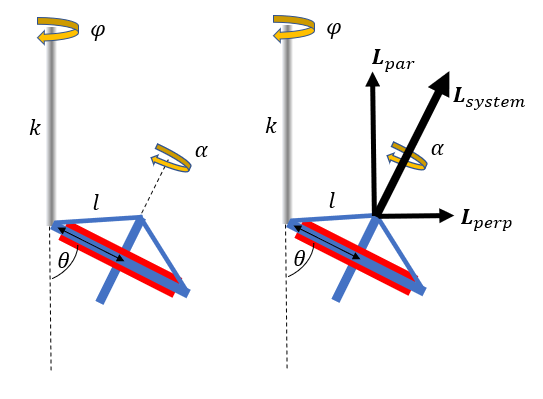}
\caption{Left: Definition of variables. Right: Decomposition of angular momentum.}
\label{scheme}
\end{figure}

\begin{figure}[ht]
\centering
  \includegraphics[width=0.9\textwidth]{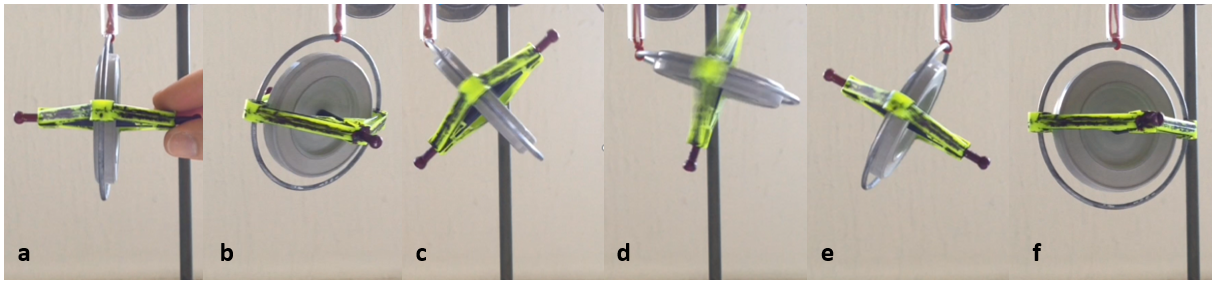}
\caption{Subsequent phases of the gyroscope motion. a) Release of the spinning gyroscope. b) Rising of inclination angle $\theta$. c) Rotation of gyroscope around the thread. d)  Inclination reaches its maximum, gyroscope spins fast around the thread. Thread is almost completely unwound. e) Thread starts to wind again. f) Inclination angle decreases back to $\theta = \SI{0}{\degree}$.   }
\label{motion}
\end{figure}

In order to observe the phenomenon, one should twist the thread and spin the gyroscope fast enough. The gyroscope is then released from its initial position when the axis of the gyroscope and thread are perpendicular ($\theta = \SI{0}{\degree}$, Fig. \ref{scheme}). Thread starts to unwind and the inclination angle $\theta$ rises, up to the moment when the gyroscope axis is almost parallel to the thread and the gyroscope spins very fast around it. Torsional resistance of the thread then causes slowing down of spinning motion around the thread and lowering of inclination angle $\theta$. One thus observes periodical oscillations in the inclination angle $\theta$ coupled with oscillations in the azimuthal angle $\phi$ around the axis of the thread (Fig. \ref{motion}).

\subsection{Minimal model}
Let us assume that the gyroscope spins fast ($\dot{\alpha} \gg \dot{\phi}, \dot{\theta}$) and its spinning angular velocity $\dot{\alpha}$ does not change in time. In other words, the major contribution to angular momentum of the whole system $\mathbf{L}_{\textit{system}}$ comes from inner angular momentum of the gyroscope, $\mathbf{L}_{\textit{system}} \approx I_3 \dot{\alpha}$, where $I_3$ is moment of inertia of the gyroscope around its axis. Relevant torques acting on the system are torsional torque of the thread $\mathbf{\tau}_{\text{thread}} = - k\phi$, where $k$ is torsional resistance of the thread, $\phi$ is azimuthal angle and $\phi = \SI{0}{\degree}$ corresponds to the state of completely unwound thread. The other relevant torque acting on the system is $\mathbf{\tau}_{\text{grav}} = mgl\sin\theta$.

By recalling the analogy with the problem of precession of heavy symmetric top, one can derive equations of motion for $\theta$ and $\phi$. The equation of motion for $\alpha$ is given implicitly under the assumption that $\dot{\alpha}$ does not change in time. Total angular momentum $\mathbf{L}_{\textit{system}}$ can be decomposed into two parts,
$\mathbf{L}_{\textit{perp}}$ perpendicular to the thread and $\mathbf{L}_{\textit{par}}$ parallel to the thread. For fixed $\theta$, magnitude of $\mathbf{L}_{\textit{perp}}$ does not change and only traces a circle around the thread, therefore revoking an analogy in analysis of heavy symmetric top, for fixed $\theta$,
$\mathrm{d}\mathbf{L}_{\textit{system}} = \mathbf{L}_{\textit{perp}} \mathrm{d}\phi =  \mathbf{L}_{\textit{system}}\cos\theta = \mathbf{\tau}_{\text{grav}}dt$.
This implies, up to the first order in $\theta$ (assuming $\theta \ll 1$):
\begin{equation}
  \dot{\phi} \approx \frac{mgl}{L_{\text{system}}} \approx \frac{mgl}{I_3\dot{\alpha}}\tan\theta \approx \frac{mgl}{I_3\dot{\alpha}}\theta\,\text{.}
\end{equation}

By repeating the same argument, now for fixed $\phi$, we end up with a conclusion that $\mathbf{L}_{\textit{system}}$ now traces a circle in $\theta$. The change of momentum
$\mathrm{d}\mathbf{L}_{\textit{system}}$ is now proportional to the component of torque $\mathbf{\tau}_{\textit{thread}}$ perpendicular to $\mathbf{L}_{\textit{system}}$.
Therefore, for fixed $\phi$, $\mathrm{d}\mathbf{L}_{\textit{system}} = \mathbf{L}_{\textit{system}} \mathrm{d}\theta  = \mathbf{\tau}_{\text{thread}}\cos\theta dt$.
This implies, up to the first order in $\theta$ and $\phi$ (assuming $\theta \ll 1$),
\begin{equation}
  \dot{\theta} \approx -\frac{k}{L_{\text{system}}}\phi\cos\theta \approx -\frac{k}{I_3\dot{\alpha}}\phi\,\text{.}
\end{equation}

These two coupled equations can be easily solved by differentiating one and inserting it into the other. Together with initial conditions $\phi(t = 0) = \phi_{\textit{max}}, \theta(t = 0)= 0$, it yields the solution:
\begin{align}
\phi (t) &= \phi_{\textit{max}}\cos\left(\sqrt{\frac{mglk}{I_3^2{\dot{\alpha}}^2}}t\right)\text{,}\\
\theta (t) &= -\phi_{\textit{max}}\sqrt{\frac{k}{mgl}}\sin\left(\sqrt{\frac{mglk}{I_3^2{\dot{\alpha}}^2}}t\right)\text{,}
\end{align}
which is valid for $\dot{\alpha} \gg \dot{\theta}, \dot{\phi}$ and $\theta \ll 1$. This minimal approach explains dynamics of torsional pendulum near $\theta \ll 1$ and gives a glimpse into explanation of the phenomenon, which consists of coupled oscillations in angles $\phi$ and $\theta$.

Up to this point, the solution is easily reproducible at advanced high school level. Approximate theory is not difficult to understand and experiments are easily performed and analysed using high-speed video recordings. Furthermore, one can introduce linear viscous damping in the thread into the theory without much additional effort. Viscous damping continuously decreases amplitude of oscillations in both angles. However, it should be noted that this kind of damping itself does not change the period of oscillations as a function of time, as can be seen in the experiment. This is why, in the next part, we introduce full equations of motion and solve them numerically.

\subsection{Full equations of motion}
In order to derive full equations of motion, we introduce the Lagrangian in terms of $(\phi, \theta, \alpha, \dot{\phi}, \dot{\theta}, \dot{\alpha})$ consisting of several terms,
where $I_2$ is moment of inertia around the axis which lies in the plane of the gyroscope disc and $I_3$ is the moment of inertia which is perpendicular to the gyroscope disc. Therefore,
\begin{align}
  L &= \underbrace{\frac{1}{2}I_2{\left({\dot{\theta}}^2+\dot{\phi}^2{\cos^2(\theta)}\right)}+\frac{1}{2}I_3{\left(\dot{\alpha}+\dot{\phi}\sin(\theta)\right)}^2}_{\text{Kinetic energy of the free gyroscope}}\\
    &+ \underbrace{\frac{1}{2}m\left(l^2\dot{\theta}^2+l^2\sin^2(\theta)\dot{\phi}^2\right)}_{\text{Kinetic energy of the center of mass}}\\
    &- \underbrace{\frac{1}{2}k{\phi}^2+mgl\cos(\theta)}_{\text{Potential energy}}\text{.}
\end{align}

Using canonical procedure, one obtains three coupled second order differential equations, from which second derivatives can be expressed as follows:
\begin{align}
  \ddot{\theta} &= -\frac{mgl\sin\theta - I_3\cos\theta\dot{\alpha}\dot{\phi}+(I_2-I_3-ml^2)\cos\theta\sin\theta{\dot{\phi}}^2}{I_2+ml^2}\\
\ddot{\phi} &= -\frac{k\cosec^2\theta\phi + \cotan\theta\dot{\theta}(I_3\cosec\theta \dot{\alpha}+(I_3-2I_2+2ml^2)\dot{\phi})}{ml^2+I_2\cotan^2\theta}\\
\ddot{\alpha} &= -\frac{I_3\cos\theta\left(I_2\cos\theta + (I_3-ml^2)\sin^2\theta\right)\dot{\theta}\dot{\phi}}{I_2I_3\cos^2\theta+I_3ml^2\sin^2\theta}\nonumber\\
		 & -\frac{I_3\sin\theta\left(-k\phi - I_3\cos\theta\dot{\alpha}\dot{\theta} +(I_2-I_3-ml^2)\sin(2\theta)\dot{\theta}\dot{\phi}\right)}{I_2I_3\cos^2\theta+I_3ml^2\sin^2\theta}
\end{align}

\begin{figure}[ht]
\centering
  \includegraphics[width=0.6\textwidth]{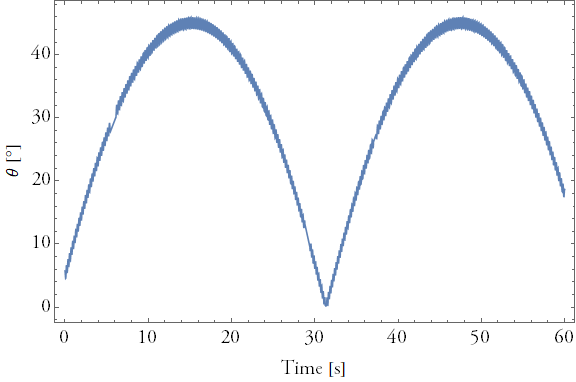}
\caption{Numerically obtained time evolution of $\theta(t)$ for 
$(\theta,\phi,\alpha)(t=0)=(0.1,80,0 \,\text{rad})$ and $\dot{\alpha}(t=0)=2\pi\cdot 5\,\text{rads}^{-1}$.
One can notice small perturbation which correspond to fast oscillations around the main evolution of the angle $\theta$ - nutation.}
\label{numerics}
\end{figure}

One can then solve these equations numerically using implicit Runge-Kutta method e.g. in Wolfram Mathematica (Fig. \ref{numerics}). Using these results, one can further describe more complex parts of gyroscope motion, e.g. nutation in $\theta$ coordinate. Using an ansatz for time dependence of $\alpha(t)$ that accounts for damping of inner spinning motion of the gyroscope causes the period of oscillations in both angles $\theta$ and $\phi$ to increase. The viscous damping of the thread then adds additional damping, which, however, again does not affect the period of oscillations.

\begin{figure}[ht]
\centering
  \includegraphics[width=0.6\textwidth]{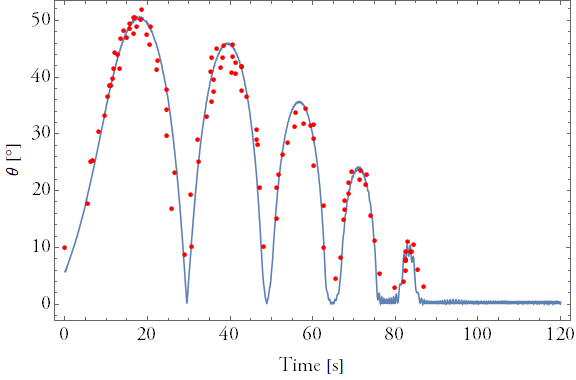}
\caption{Comparison of experimental data (red points) obtained from video analysis of gyroscope motion and numerically obtained time evolution of $\theta(t)$ for $(\theta,\phi,\alpha)(t=0)=(0.1,80,0 \,\text{rad})$ and ansatz $\dot{\alpha}(t)=\alpha_0\exp{(-ct)}$ that models the internal damping of own gyroscope spinning motion with fitted values of $\alpha_0=100\,\text{rads}^{-1}$ and $c=0.075\,\text{s}^{-1}$.}
\label{comparison}
\end{figure}

In order to compare numerical model, we performed experiment, measuring $\theta(t)$ using high-speed video analysis (see Fig. \ref{comparison}), the damping of the internal spinning motion using the ansatz for the $\dot{\alpha}(t)=\alpha_0\exp(-ct)$ with fitted damping coefficient $c$. The damping of torsional oscilations is modeled using viscous damping with coefficient $8\cdot 10^{-8}\text{s}^{-1}$. The other parameters of the gyroscope are $m = 0.15\,\text{kg}$, $l=0.045\,\text{m}$, and string of the torsional stiffness $k = 87\cdot 10^{-7}\,\text{kg}\text{m}^2\text{s}^{-2}$.

\section{Conclusion}

International Young Physicists' Tournament is an extraordinary event for high school students interested in physics and natural sciences in general. Reaching its 30\ts{th} anniversary this year, one can say it is mature enough to be a stabilised and well respected conduct attracting teams from all over the Globe, but still young and modern to adapt to quickly changing needs of nowadays society. In the first place it develops interest and talent in physics in participating students. But, equally important, it prepares them in many other competencies - research in literature, teamwork, presentation techniques, scientific discussion and judgement and many more. This is probably why, contrary to the worldwide regression of interest in STEM subjects, IYPT grows both in quality and quantity.
For 2018 a new set of $17$ problems is already available at \cite{iypt.org}. Students can decide to devote to experiments by building a seismograph, Curie point engine or Tesla valve, play around with powders changing colour or forming piles, blow bubbles, put candles in water instead of in the wind, or just trow half-filled bottles or put straws into fizzy drinks. Everyone can find a task of their interest and everyone is invited to take part in the regional rounds of IYPT competition in their countries, eventually reaching the international round next year in July in Beijing.

\section*{Acknowledgements}
This research was supported by the joint Czech-Austrian project MultiQUEST (I 3053-N27 and GF17-33780L), as well as project VEGA 2/0043/15.


\section*{References}
\begin{thebibliography}{10}

\bibitem{iypt.org}
\href{www.iypt.org}{International Young Physicists' Tournament (IYPT)}, www.iypt.org

\bibitem{IYPT2016}
M. Plesch, F. Eller, C. Kanitz, J. Landgraf, A. Raab, and S. Selbach,
\textit{The International Young Physicists' Tournament}, Eur. J. Phys. \textbf{38} (2017)

\bibitem{IPhO}
\href{http://ipho.org/}{International Physics Olympiad (IPhO)}, www.ipho.org

\bibitem{IJSO}
\href{http://www.ijsoweb.org/}{International Junior Science Olympiad (IJSO)}, www.ijsoweb.org

\bibitem{problems}
\href{http://iypt.org/images/f/f1/problems2017.pdf}{Problems for IYPT 2017}, http://iypt.org/images/f/f1/problems2017.pdf

\bibitem{vrtulka}
 M. Marek, M. Badin, and M. Plesch, \textit{Gee-Haw Whammy Diddle},
 arXiv:1707.01129v1 (2017)


\end{thebibliography}
\end{document}